\documentstyle[prl,aps,twocolumn,epsfig]{revtex}
\newcommand{\be}{\begin{equation}}
\newcommand{\ee}{\end{equation}}
\newcommand{\bdis}{\begin{displaymath}}
\newcommand{\edis}{\end{displaymath}}

\begin{document}
\bibliographystyle{prsty}  
\title{Helicity Transfer in Turbulent  Models}
\author{L. Biferale$^{1}$, D. Pierotti$^2$ and F. Toschi$^{3}$ }
\date{\today}
\address{$^{1}$  Dipartimento di Fisica, Universit\`{a} di Tor Vergata
Via della Ricerca Scientifica 1, I-00133 Roma, Italy and \\
Istituto Nazionale di Fisica della Materia, unit\`a di Tor Vergata\\
$^{2}$ Dipartimento di Fisica, Universit\`{a} dell' Aquila,
Via Vetoio 1, I-67010 Coppito, L'Aquila, Italy and\\
Istituto Nazionale di Fisica della Materia, unit\`a dell'Aquila\\
$^{3}$ Dipartimento di Fisica, Universit\`{a} di Pisa,
Piazza Torricelli 2, I-56126, Pisa, Italy and\\
Istituto Nazionale di Fisica della  Materia, unit\`a di Tor Vergata and\\
Istituto Nazionale di Fisica Nucleare, sezione di Pisa}
\maketitle
\begin{abstract}
Helicity transfer in a shell model of turbulence is investigated.
We show that a Reynolds-independent helicity flux is present in the model when
the large scale forcing 
 breaks inversion symmetry.\\
The equivalent in Shell Models of the ``2/15 law'', obtained  from helicity 
conservation in Navier-Stokes eqs., 
is derived and tested.\\
The odd part of helicity flux statistic is found to be dominated
by a few very intense events.\\
In a particular model, we calculate analytically leading and sub-leading
 contribution to the scaling of triple velocity correlation.
\end{abstract}
  
One of the most intriguing problems in three dimensional 
fully developed turbulence (FDT) 
is related to the 
appearance of anomalous scaling laws at high Reynolds numbers, i.e. 
in the limit when Navier-Stokes dynamics is dominated by the non-linear
interactions.

The celebrated 1941 Kolmogorov theory (K41) was able to capture the
main  
phenomenological ideas by performing dimensional analysis based
on the energy transfer mechanism. 
Kolmogorov   postulated that the energy cascade should
follow a  self-similar and homogeneous process entirely dependent
on the energy transfer rate, $\epsilon$. This idea, plus the assumption of 
local isotropy and universality at small scales, 
 led  to a precise prediction on the statistical
properties of the increments of turbulent velocity fields:
$\delta v (l) \sim |v(x+l)-v(x)| \sim (l\epsilon(l))^{1/3}$.  From this the
scaling of moments of $\delta v (l)$, the structure functions,
can be determined in terms of the statistics of $\epsilon(l)$, i.e.
\be
S_p(l) \equiv \langle(\delta v(l))^p\rangle=C_p 
\langle (\epsilon(l))^{p/3} \rangle l^{p/3} ,
\label{eq.1}
\ee
where $C_p$ are constants and the scale $l$ is supposed to be in 
the inertial range, i.e. much smaller than the integral scale 
and much larger than the viscous dissipation cutoff.
If $S_p(l)\sim l^{\zeta(p)}$ and
 $\langle \epsilon^p(l)\rangle\sim l^{\tau(p)}$
then
\be
\zeta(p)=p/3+\tau(p/3).
\label{eq.2}
\ee

In the K41 the $\epsilon(l)$ statistic   is assumed to  be 
$l$-independent, or $\tau(p)=0$, implying
$\zeta(p)={p\over3}, \forall p$.
On the other hand, there are many experimental and numerical
 \cite{MS,BCTBS} 
results telling  us that K41 scenario for homogeneous
and isotropic turbulence is quantitatively wrong. Strong intermittent
bursts in the energy transfer have been observed and non trivial $\tau(p)$
set of exponents measured. 

Moreover, the problem of 
investigating scaling properties of observables with the
same physical dimensions but with different tensorial structures
has been only  recently addressed \cite{procaccia1,procaccia2}.\\

Many different authors have focused their attention on the possible 
role played by helicity, the second global invariant of 3d Navier-Stokes 
eqs. \cite{kraic,lev,waleffe,bk},  for determining leading or sub-leading 
 scaling properties of correlation functions in the inertial range.

Recently \cite{procaccia2,russo}, 
an exact scaling equation for the third order
 velocity correlations entering in the  helicity flux definition 
has been derived under two hypothesis: (i) there exists
 a non-vanishing helicity flux,
(ii) the flux becomes Reynolds independent in the limit of FDT. 
This relation 
 predicts a $r^2$ scaling for a particular  third order  velocity 
correlation. 
The new relation has been called ``2/15 law'' because of the coefficient
appearing in front of the $r^2$ in analogy with the ``4/5 law''
derived by Kolmogorov for the third-order structure functions entering
in the expression of energy flux. In the "4/5" law the scaling 
 of a different third-order velocity correlation is 
found to be linear in $r$. \\
 
This simple fact tells us that different velocity correlation with the
 same physical dimension but with different tensorial structure 
may have very different scaling properties. 

Moreover, even if overwhelming evidences indicate that the main 
physics is driven by the energy transfer, there can  be 
 some sub-leading new intermittent statistics  hidden in the
 helicity flux properties. \\

Homogeneous and isotropic turbulence has, by definition, always a vanishing
mean helical flux. Nevertheless, both fluctuations about the zero-mean in
 isotropic cases and/or net non-zero fluxes in cases where inversion
 symmetry is explicitly
broken  can be of some interest for the understanding of fully developed 
turbulence. 

In this letter, we  analyse the helical transfer mechanism in dynamical models
of turbulence \cite{leo,bk,bbkt},
 built such as to explicitly  consider helicity conservation
in the inviscid limit. 

  We give the first strong numerical evidence that a 
 Reynolds-independent helicity flux is present in cases where the forcing 
 mechanism explicitly
breaks inversion symmetry. 

We confirm that in all cases
where two fluxes can coexist in the inertial range, 
velocity 
correlations
with the same physical dimension but with different transformation
properties under
inversion symmetry, can show strongly different scaling behaviour. 

In the following, we briefly  summarize the main  motivation
behind the introduction of Shell Models for turbulence. We present the 
equivalent
derivation of the ``2/15 law'' for the helicity flux in our Shell Model 
language
and we conclude with our numerical results about the Reynolds independency 
of helicity flux and about its statistical intermittent properties. 

Shell models have demonstrated to be very useful for the understanding
of many properties connected to turbulent flows \cite{G}-\cite{ls}.
The most popular shell model, the Gledzer-Ohkitani-Yamada 
(GOY) model (\cite{G}-\cite{ls}), has been shown to predict
scaling properties for $\zeta(p)$ (for a suitable  choice of the free
parameters) similar to what is found experimentally.  

The GOY model can be seen as a severe truncation of the 
Navier-Stokes equations: 
it retains only one complex mode $u_n$ as a representative 
of all Fourier modes in 
the shell of wave numbers $k$ between $k_n=k_02^n$ 
and $k_{n+1}$.

It has been pointed out that GOY model
conserves in the inviscid, unforced limit 
two quadratic quantities. The first quantity is 
the {\it{energy}}, while the second is 
the equivalent of {\it{helicity}} in 3D turbulence 
\cite{BKLMW}.  In two recent works \cite{bk,bbkt} the GOY
model has been generalized in terms of shell variable, $u_n^+, u_n^-$,
transporting positive and negative helicity, respectively.  It is easy to 
realize that only 4 independent classes of models  can be 
derived such  as to preserve the same helical structure of Navier-Stokes 
equations \cite{waleffe}.
 All  these models have 
at least one inviscid invariants non-positive defined which is
similar to the 3D Navier-Stokes helicity.  In the following, we will
focus on the intermittent properties of a mixture of two of such a models. 
The mixture is a linear combination of the old GOY model 
(extended to have $u^+,u^-$) plus another
model which has a different helical interaction and which has already been 
extensively investigated \cite{bbkt,bbt}. We focus only on two
of the four possible models because they are the only
two classes of models which show a clear forward 
energy cascade (see \cite{bbkt} for more
details). The time evolution
for positive-helicity shells reads \cite{bbkt}:
\be
\dot{u}^+_n=i k_n (A_n[u,u]+xB_n[u,u])^* -\nu k^2_n u^+_n 
+\delta_{n,n_0}f^+,
\label{eq:shells}
\ee
with the equivalent eqs, but with all helicity signs reversed, for  
$\dot{u}^-$.
In (\ref{eq:shells}),  $x$ defines the relative weights of the two models 
in the mixture, 
$\nu$ is the molecular viscosity, $f^+, f^-$ are the large scale forcing and  
$A[u,u]$ and $B[u,u]$ refer to the non-linear terms of the two models. Namely:
\begin{eqnarray}
A_n[u,u] &\equiv &u^{-}_{n+2} u^{+}_{n+1}+b_3
u^{-}_{n+1} u^{+}_{n-1}
+c_3 u^{-}_{n-1} u^{-}_{n-2},\\
B_n[u,u] &\equiv &u^{+}_{n+2} u^{-}_{n+1}+b_1 
u^{-}_{n+1} u^{-}_{n-1}
+c_1 u^{-}_{n-1} u^{+}_{n-2}.
\end{eqnarray}

It is easy to verify that for the choice $b_3=-5/12,c_3=-1/24,
b_1=-1/4,c_1=-1/8$ there are only two global inviscid invariants
\cite{bbkt}:
 the energy, $E= \sum_n(\vert u^+_n \vert^2 +
 \vert u^-_n \vert^2))$,  and helicity, $H=\sum_n k_n
(\vert u^+_n \vert^2 - \vert u^-_n \vert^2)$.

The equations for the fluxes throughout shell number $n$ are:
\begin{eqnarray}
d/dt\sum_{i=1,n}E_i= k_n <(uuu)_n^E>-\nu\sum_{i=1,n}E_i +  E_{in},
\label{flussi0}\\
d/dt\sum_{i=1,n}H_i= k_n^2<(uuu)_n^H>  -\nu\sum_{i=1,n}H_i+ H_{in},
\label{flussi}
\end{eqnarray}
where $E_i$ and $H_i$ are  the  energy and helicity of the $n$th shell,
respectively:
 $E_i= <|u_i^+|^2 +|u_i^-|^2 >$, $H_i =k_i(<|u_i^+|^2 -
|u_i^-|^2 >)$.  $E_{in}$ and $H_{in}$ 
are the input of energy and helicity due to forcing effects,
$E_{in}= \Re(<f^+(u^+_1)^{*}+f^-(u^-_1)^{*}> )$, 
$H_{in}=\Re(k_1<f^+(u^+_1)^{*}-f^-(u^-_1)^{*}> )$.\\ 
In (\ref{flussi0}) and (\ref{flussi})  
we have introduced the triple correlation:
\begin{eqnarray}
<(uuu)_n^E> &= &(\Delta^+_{n+1}+\Delta^-_{n+1})+
(b_3+1/2)(\Delta^+_{n}+\Delta^-_{n})
\nonumber\\
        &+&  x \left((\Gamma^+_{n+1}+\Gamma^-_{n+1})+
(b_1+1/2) (\Gamma^+_{n}+\Gamma^-_{n})\right),
\label{triple0}\\
<(uuu)_n^H> &=& (\Delta^+_{n+1}-\Delta^-_{n+1})+
(b_3+1/4) (\Delta^+_{n}-\Delta^-_{n}) \nonumber \\
        &+&  x \left((\Gamma^+_{n+1}-\Gamma^-_{n+1})+
(b_1+1/4) (\Gamma^+_{n}-\Gamma^-_{n})\right),
\label{triple} 
 \end{eqnarray}
and 
\be
\Delta_n^+= \left\langle \Im(u^+_{n+1}u^-_nu^+_{n-1})\right\rangle, \;\;\; 
\Gamma_n^+= \left\langle \Im(u^-_{n+2}u^+_{n+1}u^+_{n})\right\rangle .
\label{deltagamma}
\ee
Assuming 
that there exists a stationary state  we have 
${d\over dt} \Pi^{E}_{n}={d \over dt} \Pi^{H}_{n}=0$,
where $\Pi_{n}^{E}=k_{n}\langle (uuu)_{n}^{E}\rangle$ and
$\Pi_{n}^{H}=k_{n}^{2}\langle (uuu)_{n}^{H}\rangle$. Moreover,
in the inertial range we can neglect the viscous contribution 
in (\ref{flussi0}) and (\ref{flussi}), obtaining:
\be
<(uuu)_n^E>=k_n^{-1}E_{in}, \;\;\;
<(uuu)_n^H>=k_n^{-2}H_{in}.
\label{45}
\ee
Supposing that there exist the  energy and helicity
 fluxes, $E_{in}=\epsilon$, $H_{in}=h$ (the latter different from zero 
 only if $f^+ \neq f^-$) and supposing that both are Reynolds-independent,
we have in the inertial range:
\be
<(uuu)_n^E> \sim k_n^{-1},\;\;\;
<(uuu)_n^H> \sim k_n^{-2}.
\label{scaling}
\ee
Relation (\ref{45}) is the equivalent of what found for helical Navier-Stokes 
turbulence in \cite{procaccia2,russo}).

 Figure 1 reports results for the helicity flux in a numerical simulations
done with two different Reynolds numbers, $Re\sim 10^{5}$, 
$Re\sim 10^{9}$
for a choice of mixture parameter $x=0.1$ and other numerical inputs
as follows, $N=16$ and 26, $\nu=10^{-5}$ and $2 \cdot 10^{-9}$. 
A non-zero helical flux was
obtained using a forcing term breaking inversion symmetry:
$f^+=5\cdot 10^{-3}(1+i)$, $f^-=f^{+}/10$. Time
marching was obtained by using a slaved Adams-Bashforth algorithm for
a number of iterations equal to several thousands of the typical eddy turnover
time. 
A clear inertial range with a 
non-vanishing helicity flux is detected. The extension of the range
where helicity flux is roughly constant  scales
with the Reynolds number. Moreover, the flux
intensity is roughly constant at changing Reynolds number,
giving the first evidence that the model
can simultaneously support both  energy and helicity transfers
and that both of them are Reynolds-independent.
 
Let us remark that this is only possible due to the non-positiveness
 of helicity; 
in 2d turbulence, for example,  
a similar results, concerning enstrophy and energy cascades,
 is clearly apriori forbidden. 
 
 As for the statistics of helicity transfer, we measured the scaling exponents
of the moments of energy and helicity fluxes:
\begin{eqnarray}
\Sigma_{E}^{(p)}\equiv<((uuu)_n^E)^{(p/3)}> \sim k_n^{-\zeta(p)}\\
\Sigma_{H}^{(p)}\equiv<((uuu)_n^H)^{(p/3)}> \sim k_n^{-\psi(p)}.
\label{moments}
\end{eqnarray}
As one can see in figure 2 we have that
the even part of the two statistics coincides, i.e. $\zeta(2p)=\psi(2p)$.
On the other hand, the  scaling exponents of odd moments are different.

The difference in odd moments is the signature of 
strong cancellation effects in the statistics connecting fluctuations
at different scales. The picture we have in mind is that the main 
effect driving turbulent fluctuations is due to the
energy cascade process, with its intermittent fluctuations
measured by the $\zeta(p)$ exponents. 
Superimposed to the energy transfer, there
are "topological" fluctuations introduced by the asymmetric forcing
and measured by the odd-part of the helicity turbulent transfer.

Let us notice that helicity flux fluctuations are much larger then the 
average helicity flux. This clearly distinguishes the helicity transfer 
mechanism from the energy transfer mechanism. \\
In the former, the strong
intermittent behaviour shown by odd moments tell us that the odd part of the 
statistic is dominated by a few very singular structures.

In the case of no-mixture ($x=0$) one can also exactly calculate  
sub-leading scaling
for the triple correlation $\delta^{+}_n \equiv k_n(\Delta_n^++\Delta_n^-),\,
\delta^-_n\equiv k_n^2(\Delta_n^+-\Delta_n^-)$.
Indeed, from expressions (\ref{triple0})
one obtains after some simple algebra:
\begin{equation}
\delta^{+}_n = 2\epsilon \frac{1-y^{n+1}}{1-y}, \;\;\;
 \delta^{-}_n = 4h \frac{1-z^{n+1}}{1-z}
 \end{equation}
 where $y=-2(b_3+1/2)$ and $z=-4(b_3+1/4)$. Being both $y$ and $z$ 
 with modulus less then one we recover the asymptotic scaling 
 (\ref{scaling}) and one can also control sub-leading correction to it:
$$\Delta_n^+ = k_n^{-1}(\epsilon \frac{1-y^{n+1}}{1-y}) +k_n^{-2}
(2h \frac{1-z^{n+1}}{1-z})$$ and
$$\Delta_n^- = k_n^{-1}(\epsilon \frac{1-y^{n+1}}{1-y}) -k_n^{-2}
(2h \frac{1-z^{n+1}}{1-z})$$

 In conclusion, we have studied a Helical shell model for turbulence
 with a forcing which explicitly breaks  inversion symmetry at 
 large scale. \\
We have checked that a Reynolds-independent helicity flux
 establishes  in the system, giving the first evidence of
 very different scaling for triple correlations entering 
 in the energy flux and helicity flux definitions.\\ 
The odd part of the helicity flux  statistic
 is found to be strongly intermittent.\\
  For a particular class of models
 we can also calculate explicitly sub-leading corrections to 
 pure scaling behaviour of typical triple correlation functions.\\
 The existence of sub-leading terms explicitly tell us that
 scaling laws in turbulent flows must be studied 
 on correlation functions which have a pure 
 projection on the physical relevant quantities. \\
There are not reasons why similar effects should not be present
also in true Navier-Stokes eqs. For example,
  spurious intermittent corrections can be detected
in cases where isotropy is globally or locally violated (as in
boundary layers).
 
\noindent
{\bf Acknowledgements}: This work has been partially supported by the INFM
(Progetto di Ricerca Avanzata: TURBO).

\centerline{FIGURE CAPTIONS}
\begin{itemize}
\item Figure 1:\\ 
Helicity-flux, $(k_{n}^{2}<(uuu)_{n}^{H}>)$, versus $k_{n}$ 
for $N=16$, $\nu=10^{-5}$ (dashed line) and $N=26$, $\nu=2\cdot 10^{-9}$
(continuos line). 
 
\item Figure 2:\\ 
Anomalous exponents for the helicity flux, $\psi_{p}$ (circles), and
for  the energy flux, $\zeta_{p}$ (squares), obtained for $N=26$ and 
$\nu=2\cdot 10^{-9}$.
\end{itemize}

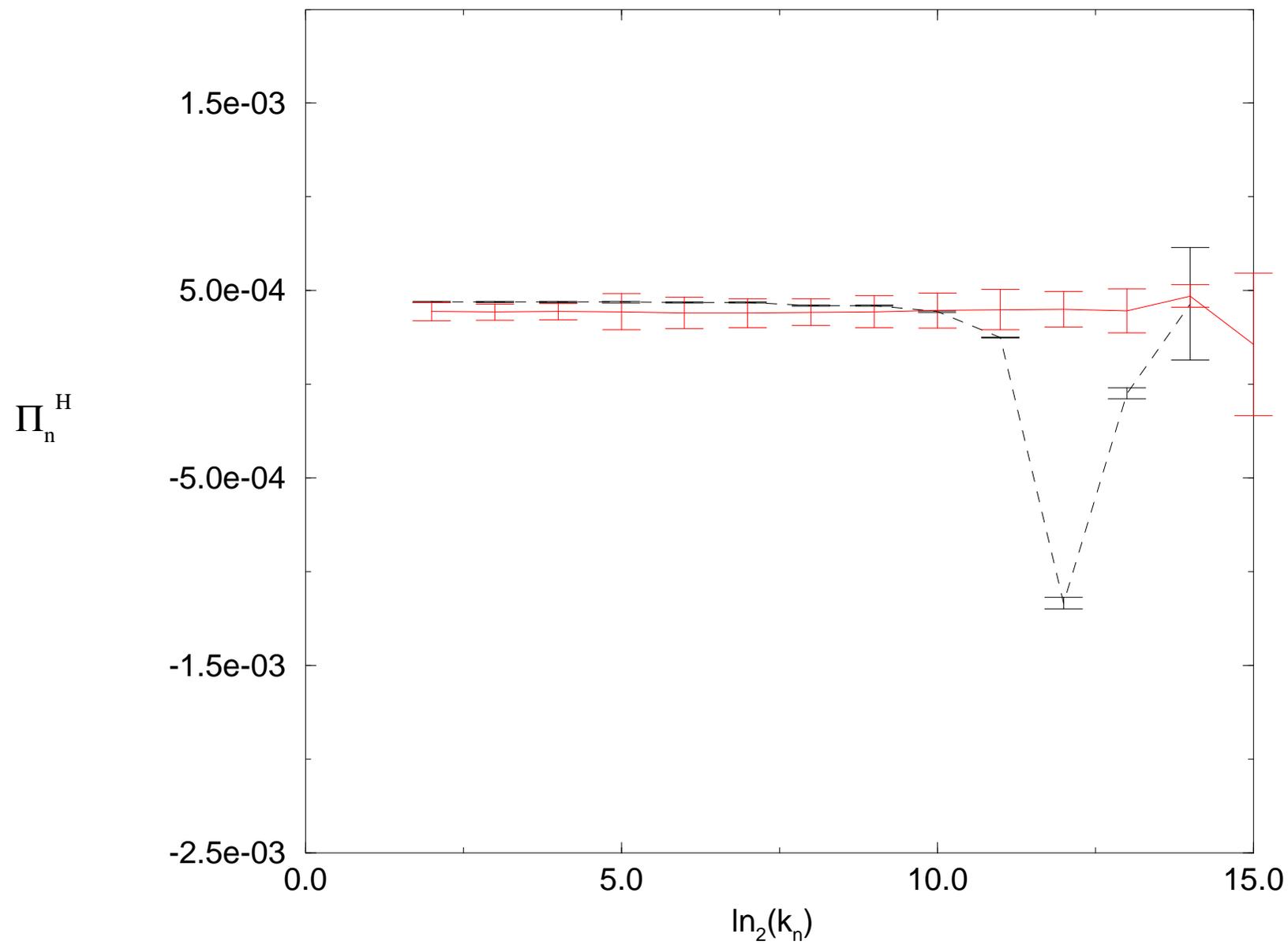
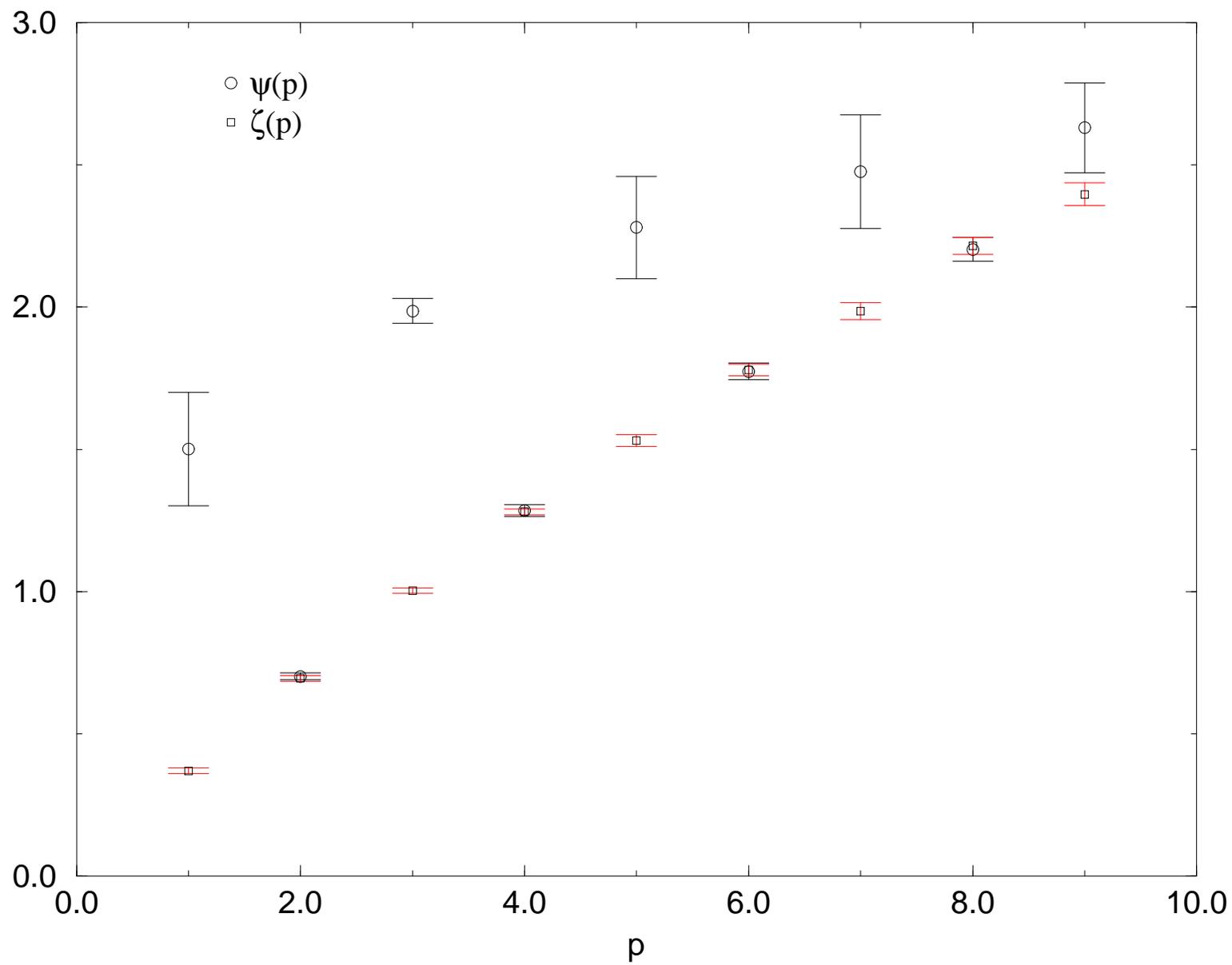
\end{document}